\documentclass[conference,10pt]{IEEEtran}
\usepackage[final]{graphicx}
\usepackage{mathrsfs,amsmath}
\usepackage[outdir=./]{epstopdf}
\usepackage{color,epsfig,calc}
\usepackage{todonotes}
\usepackage{amssymb}
\usepackage{multirow} 
\PassOptionsToPackage{hyphens}{url}\usepackage{hyperref}
\usepackage{balance}

\definecolor{lightgreen}{RGB}{170, 245, 140}

\title{Multi-User Frequency-Selective Hybrid MIMO Demonstrated Using 60 GHz RF Modules}
\author{Steve Blandino$^{1,2}$, Claude Desset$^{2}$,  Cheng-Ming Chen$^{1}$,  Andre Bourdoux$^{2}$, Sofie Pollin$^{1,2}$\\
$^1$~KUL ESAT, Kasteelpark Arenberg 10, B-3001 Leuven, Belgium\\
$^2$~Imec, Kapeldreef 75, B-3001 Leuven, Belgium\\
}

    \makeatletter 
    \def\ps@IEEEtitlepagestyle{%
        \def\@oddfoot{\hfill \textcopyright 2018 IEEE \hfill}%
        \def\@evenfoot{}%
    } 
    \makeatother 

\begin{document}

\maketitle

\begin{abstract}
Given the high throughput requirement for 5G, merging millimeter wave technologies and multi-user MIMO seems a very promising strategy. As hardware limitations impede to realize a full digital architecture, hybrid MIMO architectures using both analog and digital precoding are considered a feasible solution to implement multi-user MIMO at millimeter wave. 
Real channel propagation and hardware non-idealities  degrade the performance of such systems thus experimenting the new architecture is crucial to  support system design. Nevertheless, hybrid MIMO systems are not yet understood as the effects of the wide channel bandwidths at millimeter wave, the non-ideal RF front end as well as the imperfections of the analog beamforming using phased antenna arrays are often neglected. In this paper, we present a 60 GHz multi-user MIMO testbed using phased antenna arrays at both transmitter and receivers. The base station equipped with a 32 phased antenna array allocates simultaneously two users. We show that frequency selective hybrid precoding can efficiently suppress inter-user interference enabling spatial multiplexing in interference limited scenario doubling the throughput compared to a SISO scenario and compensating the frequency fluctuation of the channel. In addition, we report an EVM constellation improvement of 6\'dB when comparing the hybrid MIMO architecture with a fully analog architecture.
\end{abstract}

\section{Introduction}
\label{sec_intro}
\IEEEpubid{0000--0000/00\$00.00˜\copyright˜2015 IEEE}


The increasing growth of the use of mobile devices and wireless services is expected to continue, and anticipated both by the research and industry community. Related to that, the
5G race is going through a crucial phase, and academia and leading companies are developing solutions and prototypes to solve the challenges of the new generation system.
After a preliminary study phase, researchers consider millimeter wave (mm-wave) and Massive MIMO as the breakthrough technologies able to guarantee the expected system capacity increase, while meeting the low power consumption, high reliability and low latency requirements. 

The available spectrum for conventional communication systems, e.g. WiFi and 4G (or older) cellular systems, which operate at carrier frequencies below 6 GHz is already close to saturation.
 The immense available spectrum compared to the sub 6\,GHz bandwidth and the rapid progress in semiconductor technology motivate the shift to  mm-wave  and in the last years several effort has been made to demonstrate the feasibility of future cellular systems operating at  28\,GHz, 39\,GHz,and 72\,GHz. Several standards operating at 60 GHz have been also released, e.g. IEEE 802.11\,ad and 802.15.3\,c, mainly for indoor networks and IEEE 802.11\,ay is expected soon. Research and industry have joined their forces to offer broadband access at mm-wave  before 2020 \cite{miwaves}\cite{miweba}\cite{mmmagic}\cite{5GCH}.
In contrast to mm-wave technologies, MIMO techniques increase the spectral efficiency by exploiting the spatial  dimension of the channel, sending simultaneously multiple streams on a single time-frequency resource. In Massive MIMO, the base station (BS)
sends these streams  to different user equipment devices (UEs) using a large number of  antenna elements to exploit   favorable propagation condition \cite{6951994}. Massive MIMO technology, using simple linear processing at the base station, averages out small-scale fading, noise and interference,  and increases rate and diversity gains.

Given the peak throughput requirement for 5G (20 Gbps as proposed by the ITU), the most promising strategy is  to combine both the high bandwidth available at mm-wave frequencies, and the spectral efficiency improvement achieved by exploiting the spatial degrees of freedom. As mm-wave enables very directive beams, even with relatively small antennas, constructing solutions with a large number of antennas and beams is a natural evolution.
In addition, larger antenna arrays can overcome the high signal attenuation at mm-wave caused by atmospheric gases.
The METIS-II project proposed mm-wave solutions with up to 8 spatial streams \cite{metis}, showing that large industry consortia also believe in the relevance of spatial multiplexing combined with mm-wave \cite{6894453}.
Mm-wave Massive MIMO, however, is not a simple shift of  the traditional Massive MIMO  to higher frequency. It is a novel  technology based on different propagation characteristics and  hardware constraints caused by the high frequency and very wide bandwidth. Mm-wave solutions for some fixed wireless access applications, where multiple beams can be constructed pointing in different directions seem feasible today. However, the realization of full mm-wave massive MIMO, creating flexible pipes  of data that adaptively changes beam patterns adapting to a large bandwidth in the frequency  domain  towards multiple, possible moving, UEs is still not yet achieved. 
From a hardware perspective, transceiver imperfections are larger at mm-waves and with wide channel bandwidths. Moreover, RF chains need to be closely packed near the antenna elements to avoid  signal propagation over long paths introducing high signal losses, coupling and distortion. However, it may be difficult to integrate many of RF chains close to the antenna elements as half-wavelength antenna spacing is generally used to avoid  grating lobes.
Finally,  power consumption is another limitation as RF devices at mm-wave are particularly power-hungry. 
These hardware limitations impede a full-digital architecture as originally proposed for the sub 6\,GHz Massive MIMO. Also, traditional fully analog mm-wave base station architectures are not suitable to support wideband multi-UE systems for two main reasons. First, fully analog systems, as prosed in the 802.11\,ad, consist of a single RF chain, thus multiple streams cannot be supported. Second, the analog beamforming is frequency flat and cannot be adaptive to the channel frequency variations which might be significant in wide channels, especially when considering dynamic scenarios. Hence, hardware constraints have led to propose hybrid analog-digital architectures as a feasible way to implement mm-wave MU-MIMO.


Experimenting multi-UE hybrid architectures at mm-wave (hybrid MIMO) is necessary to  support
system design and confirm whether such a system performs as theoretical predictions.  While MIMO is in general well understood, a hybrid MIMO system at mm-wave requires the understanding of the  mm-wave channel, including also the complex effect of the hardware non-idealities and the analog beamforming. 
Most of the practical validation of a multi-UE hybrid MIMO at  mm-wave  differs significantly  from what is expected.
Recently, several organizations have worked on prototype demonstration. 
Leading companies have proposed several 5G mm-wave solutions at 28\,GHz and a complete overview is given in \cite{inoue2016special} and \cite{sakaguchi2017and}.  However most of these studies do not use spatial multiplexing techniques to serve multiple UEs.
In the context of multi-UE scenarios,
\cite{7063460} as well as \cite{7848964} have proposed prototypes based on lens antennas which natural application is wireless
backhaul.
\cite{6866623} presents a  mm-wave  short range
communication system, which includes RF phased array front-end at 60\,GHz in which two UEs are spatially separated just by different analog beams.  
Recently,   \cite{7794711} has shown the realization of hybrid MIMO using  an interleaved antenna array consisting of 32 antenna elements grouped in  two set of 16-element antennas array  operating at 60\, GHz. The authors  confirm the beam multiplexing performance in experiments.  However, in this implementation, the digital beam weights are computed using beam index feedback from the UEs. The weights thus are not adapted to the frequency variation of the channel  and to hardware imperfections. The recent demonstration of multi-UE hybrid MIMO   \cite{NI}  used  receiver horn antennas,  making hard to include the true physical channel as they may not be representative of the often imperfect beam generated by a phased antenna array as demonstrated in \cite{saha2017x60}. Moreover, in \cite{NI} digital beamforming was not adapted to the frequency dimension. Finally, also the authors in \cite{ghasempour2017decoupling} implemented multi-UE testbed at 60\,GHz using mechanically steerable horn antennas and frequency flat digital precoding.
All existing experimental work is hence limited either to narrowband scenario or using antennas that are more suitable for fixed point-to-point backhaul applications.

    In this paper, our aim is to present a first realization of a hybrid MIMO  testbed which include phased antenna arrays at both transmitter and receiver operating at 60\,GHz. Using multiple RF chains, the BS can precode the  baseband streams using  frequency selective digital beamforming designed from  the multi-UE MIMO channel estimation, including hardware non-idealities and a realistic indoor environment.    
We compare the  performance of this hybrid architecture to an analog-only architecture in which different receivers are also simultaneously allocated in the same time-frequency resource  but spatially separated by different analog beams. 
We show that in an UE-interference limited scenario, hybrid MIMO enable spatial multiplexing.

The reminder of the paper is organized as follow: Section II introduces different architectures which can be implemented at mm-wave and the system model. Section III explains the hybrid MIMO operations. Section IV describes the testbed implementation. Section V reports measurements and Section VI gives the conclusion and an overview of future works.

\section{Hybrid Multi-User Beamforming Architectures and  System Model}
\label{system_model}
\begin{figure}[]
\centering
\includegraphics[width=0.48\textwidth]{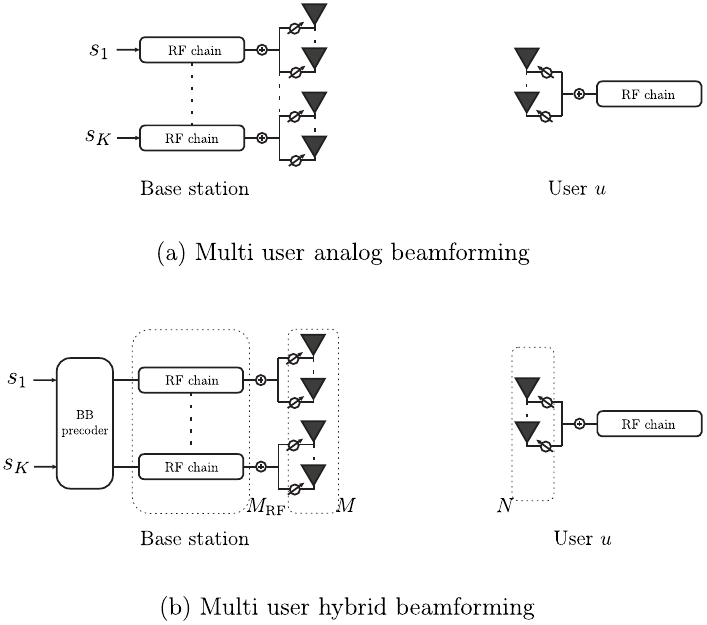}
 \caption{Analog beamforming and hybrid beamforming are two different architectures able to support multi-UE transmission. Baseband processing in hybrid beamforming is used to improve the SINR in interference limited scenarios.\label{fig:architectures}}
\end{figure}

In this section we first introduce the multi UE architectures which can be implemented at mm-wave and then we describe the system model of this paper.

We consider a BS with multiple RF chains to support multiple streams simultaneously.
As a full digital architecture is not realistic, 
Figure \ref{fig:architectures} shows two architectures which can support multi UE transmission.
Figure \ref{fig:architectures}a illustrates the full analog architecture where  each RF chain is connected to a portion of the array, here referred as sub-array. Each sub-array can send one stream of data to a single UE. 
This architecture operates assuming that the high directivity antenna pattern at 60\,GHz yields to negligible interference between adjacent transmissions.
Hybrid architecture, shown in Figure \ref{fig:architectures}b, allows to relax this assumption.
This architecture uses a combination of analog beamforming with digital beamforming. Digital baseband precoding can be designed to reduce inter user interference.

In this paper, we consider a multi-UE  mm-wave  hybrid beamforming architecture. The BS is equipped  with $M$ antennas equally distributed  in $M_\text{RF}$ sub-arrays. Each sub-array is connected to a single RF chain, hence the system consists of   $M_\text{RF}$ RF chains. This hybrid beamforming architecture is usually referred in the literature as a partially connected architecture as each RF chain can be connected only to a subset $M_\mathrm{sub}=M/M_{\text{RF}}$ of antenna elements \cite{8030501}.
The antennas are disposed in a rectangular array to perform beamforming in azimuth and elevation.   We assume to  transmit $K\le M_\text{RF}$ streams towards $K$  UEs. 
Each UE has analog beam steering capabilities since the use of $N$ antenna elements disposed in a rectangular array, connected to one single RF chain.
We consider a wideband  single carrier system.
At the transmitter the data symbols are generated in time domain and transformed in frequency domain to apply per-subcarrier precoding. The precoded symbols in frequency domain are then transformed back to the time domain. A cyclic prefix is then added before analog beamforming.
The  baseband symbol at each subcarrier $k$  in frequency domain can be written as:
\begin{equation}
\mathbf{x}[k] = \mathbf{F}_\text{A} \mathbf{F}_\text{D} [k] \mathbf{s}[k],
\end{equation}
where $\mathbf{s}[k] \in \mathbb {C}^{K\times 1}$ are the data symbols at the subcarrier $k$ such that $E{[\mathbf{s}^H\mathbf{s}]}=1$, while $\mathbf{x}[k] \in \mathbb {C}^{M\times 1} $ are the precoded symbols transmitted over the air.   
The symbols are precoded in digital domain using the frequency selective precoding matrix  $\mathbf{F}_\text{D}\in \mathbb {C}^{M_\mathrm{RF}\times K}$. The  analog precoding is  implemented using a bench of phase shifters which are represented by  $\mathbf{F}_\text{A} \in \mathbb {C}^{M\times M_\mathrm{RF}}$.
We  emphasize that the analog beamforming matrix F$_A$ is constant over the whole bandwidth. This means that the analog beamforming is  frequency flat while the baseband precoders can be different for each subcarrier and compensate the  channel fluctuations caused by multi-path propagation, particularly relevant in an indoor environment or coming from  hardware non idealities.
Considering the circulant property introduced by using cyclic prefix, the received signal at the UE $u$ is:
\begin{equation}
\mathbf{y}[k]  = \alpha \mathbf{w}_\text{A}^u\mathbf{H}^u[k] \mathbf{x}[k],
\end{equation}
where $\alpha = \frac{1}{||\mathbf{F}_{\text{A}} \mathbf{F}_{\text{D}}||}$ sets the total power constraint, $\mathbf{H}^u \in \mathbb {C}^{N\times M}$ is downlink channel observed by the UE $u$, and   $\mathbf{w}_\text{A}^u \in \mathbb {C}^{ 1\times N}$ is the receiver analog combiner vector.




\section{Hybrid Precoding}
In this section, we introduce the design of the matrices $\mathbf{F}_{\text{A}}$ and $\mathbf{F}_{\text{D}} $.
The design of these precoding  matrices has been extensively studied in the literature e.g. \cite{alkhateeb2015limited}.  
In general, two main procedures can be applied.  A first possibility is to   jointly design  $\mathbf{F}_{\text{A}}$ and $\mathbf{F}_{\text{D}} $ such that their product is as close as possible to the full digital optimal precoding. These schemes are not easy to implement in practice as they require full channel state information at each antenna to compute the optimal precoder matrix and they also require many iterations before converging to the solution.   In our testbed instead we decouple the design $\mathbf{F}_{\text{A}}$ and $\mathbf{F}_{\text{D}} $ in two stages. In the initial phase, the BS and  each UE select the optimal beam pair increasing the average received SNR in each point-to-point link, without any concern on possible interference from the other spatial streams.   After fixing  the analog beams  the BS   further optimizes the transmission  by refining the beams through digital precoding  to suppress multi-UE interference. Digital  precoding is applied on the reduced channel, which comprises analog beamforming, front-end response and the propagation channel.
This   scheme is more suitable for practical implementation as 
it assumes that the SNR changes slowly compared to the coherence time of the channel and the design of the analog matrix $\mathbf{F}_A$ does not require frequent reconfiguration. The BS requires only instantaneous channel knowledge  of the reduced channel.

\subsection{Analog Beamforming}


Let us consider a uniform planar array  lying on the plane $yz$ and broadside direction along the $x$ axis, such that $M =M_y\cdot M_z$ with equidistant half-wavelength element spacing.
In the far field zone the transmitted wave can be represented as a plane wave and the total transmitted energy is shaped by the array factor AF:
\begin{equation}
\text{AF} = \sum_{n=0}^{M_y-1} 
\left[  \sum_{m=0}^{M_z-1} 
\mathrm{e}^{\mathrm{j}m k_z}\mathrm{e}^{\beta_z(m,\theta_0)}
\right]
\mathrm{e}^{\mathrm{j}n k_y}\mathrm{e}^{\beta_y(n,\theta_0, \phi_0)} ,
\end{equation}
where $ k_y = \pi\mathrm{sin}(\theta)\mathrm{sin(\phi)}$ and  $ k_z =\pi\mathrm{cos}(\theta)$ with $\theta$ and $\phi$ being azimuth and elevation angles respectively.
The phases $\beta_z$ and $\beta_y$ are independent of each other and they represent the difference in phase excitation between the antenna elements.  They can be adjusted to control  the desired direction of the total field of the array.
Given $(\theta_0, \phi_0)$, the desired direction the phase at the antenna $(n,m)$ can be written as:
\begin{equation}
\label{eq:beta_z}
\beta_z(m,\theta_0) =-\mathrm{j} k_z m \mathrm{cos}(\theta_0),
\end{equation}
\begin{equation}
\label{eq:beta_y}
\beta_y(n,\theta_0,\phi_0) =-\mathrm{j} k_y n\mathrm{sin} (\theta_0)\mathrm{sin(\phi_0)} .
\end{equation}
%

In the partially connected architecture, each  sub-array radiation pattern is designed independently by adjusting the phase excitation of each antenna element using (\ref{eq:beta_z}) and (\ref{eq:beta_y}).  Hence, the analog beamforming  matrix is:
\begin{equation}
\mathbf{F}_\text{A}
= 
\left[
\begin{array}{cccc}
\mathbf{f}_1\\
 &\mathbf{f}_2\\
& & \ddots\\
  &  & & \mathbf{f}_{M_{\mathrm{RF}}}
\end{array} 
\right],
\end{equation}
where  $\mathbf{f}_i(\theta_0, \phi_0)\in \mathbb {C}^{M_\mathrm{sub}\times 1},\ i = 1\dots M_\mathrm{RF}$ collects the phase shifts of the antenna elements of the sub-array $i$. Each sub-array $i$ can design 
a different analog beam by setting a different $(\theta_0, \phi_0)$.
The same theory holds for the design of the receiver analog vector  $\mathbf{w}_\text{A}^u$ at the UE $u$.

To acquire the phases in (\ref{eq:beta_z}) and (\ref{eq:beta_y}) the system runs a  beam search procedure. In our testbed, during this stage the BS and sequentially  each UE scan all the beam space to select the beams pair which maximize the received SNR. Full space search is not optimal in terms of complexity and required time to find the optimal pair. However,  here we assume a quasi-static environment assuming the angles $(\theta_0, \phi_0)$  constant. The fast acquisition of the analog beamforming matrix and the tracking of these angles  in case of mobility  are  topics of great interest in mm-wave, but  out of the scope of this work.

\subsection{Digital Beamforming and reduced channel estimation}
At the BS, each sub-array has its own steering capability and the analog processing is able to create beam multiplexing by optimizing the transmitted power in target directions.
However a full analog system, is generally interference-limited and nulling interference is required. 
Instead the digital  beamforming  can produce a null along the undesired direction with simple linear precoding. Moreover, digital beamforming can be adapted to equalize channel variations along the frequency domain which cannot be done with simple analog precoding.  
 Zero-forcing (ZF) beamforming e.g. is a  practical scheme which suppress interference  while   equalizing the channel of each UE by multiplying at the transmitter the data with the channel inverse at each subcarrier.
 This scheme however might yield poor performance if the channel is badly conditioned in some subcarrier. A regularised zero-forcing (RZF) scheme improves the  ZF  performance in the subcarriers subject to  low SNR.
 Hence we design $\mathbf{F}_\text{D}$ as:
\begin{equation}
\mathbf{F}_\text{D}[k]= \tilde{\mathbf{H}}[k]^H \cdot (\gamma \mathbf{I}_K +\tilde{\mathbf{H}}[k]\tilde{\mathbf{H}}[k]^H)^{-1},
\end{equation}
where $\tilde{\mathbf{H}}[k]$ is the  $K\times M_\text{RF}$ reduced digital channel which  includes analog beamforming and the front-end response and $\gamma$ is the regularizing parameter.

The design of the digital beamforming matrix $\mathbf{F}_\text{D}$ is based on the knowledge of the channel $\tilde{\mathbf{H}}$ at the BS. As in 802.11ad, channel estimation is performed  training the downlink effective channel with Golay sequences. 
$M_\mathrm{RF}$ orthogonal Golay sequences are sent over the entire bandwidth each from a different sub-array. Each UE using the correlation properties of the Golay sequences can simultaneously estimate the downlink channels  from every sub-array without being affected by interference.
 The full estimated channel is sent back to the BS which can use the full reduced-channel knowledge to design the digital beamforming matrix.


\section{Testbed}

The testbed implemented allows multi-UE downlink wireless transmission in real-time while further processing of the received signals is performed offline. 
The system is based on the single-carrier frequency domain equalizer (SC-FDE) version of the 802.11ad standard.
A transmission bandwidth of 1.76\,GHz centered  at the carrier frequency $f=58.32$\,GHz is considered. 
The standard has been extended to support multi-UE communication. Frequency dependent digital precoding has been applied using  FFT/IFFT of size 512.
The  symbols of the header and data are generated in time domain and they are grouped into blocks of 512 symbols.  A  cyclic prefix of 128 symbols is added  as guard interval to form a complete block of 640 symbols.  
\begin{figure}
\centering
\includegraphics[width=0.48\textwidth]{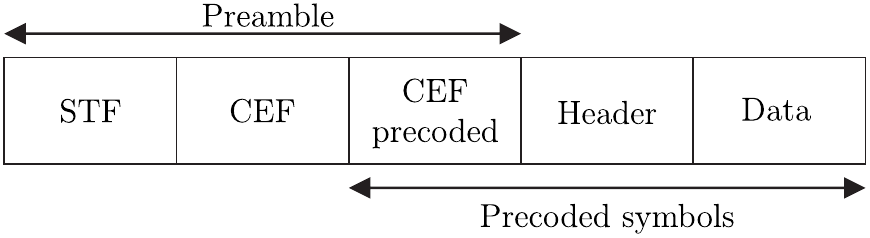}
 \caption{ A modified version of the 802.11ad frame is used to support multi-UE operation. Precoded preamble is added to perform equalization at each UE.\label{fig:frame}}
\end{figure}
The transmitted frame, shown in Figure  \ref{fig:frame}, contains a preamble at the beginning of the frame for frequency offset estimation, synchronization and channel estimation.  
The preamble includes a non-precoded and a precoded portion. The non-precoded portion is broadcasted to all the UEs. The second portion of the preamble includes the precoded CEF which allows the estimation of the precoded channel to perform frequency equalization on the precoded data. 

\begin{figure}
\centering
\includegraphics[width=0.48\textwidth]{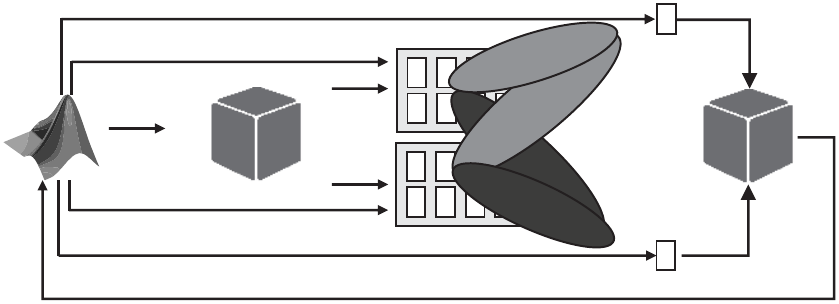}
 \caption{Partially connected hybrid BS architecture. The  PC running matlab controls BEEcube and RF front-end.    Two UEs are spatially multiplexed. \label{fig:system}}
\end{figure}
\begin{figure}
\includegraphics[width=0.48\textwidth]{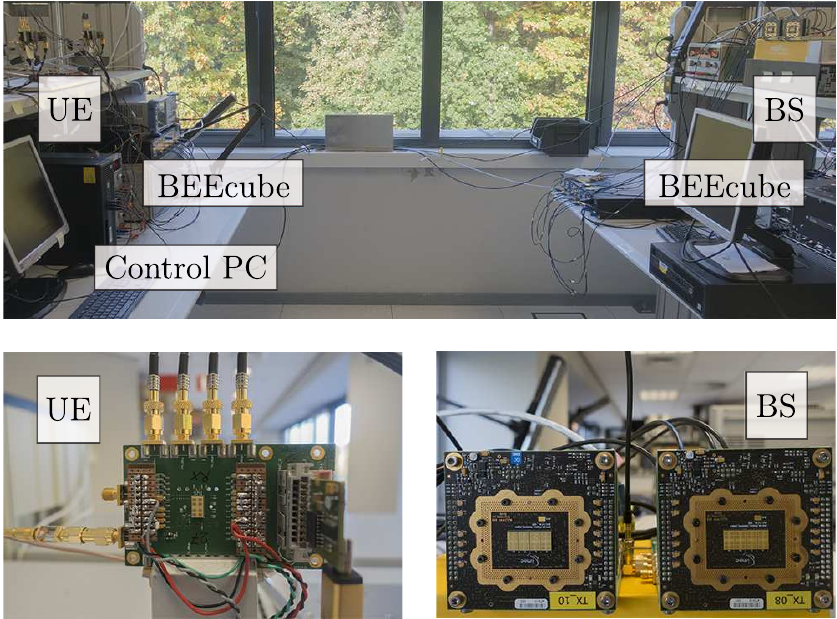}
  \caption{Top: 60\,GHz MU-MIMO testbed setup. Bottom left: UE receiver using a 2$\times$2 phased antenna array.  Bottom right: 32-antennas BS partially connected. Each RF chain is connected to a 2$\times$8 phased antenna array.\label{fig:testbed_picture}}
\end{figure}
The testbed is divided in three main sub-systems: the control PC, BEEcube platforms and mm-wave radios. An  overview is shown in Figure \ref{fig:system}. 
The control PC running Matlab, performs offline signal processing on the transmitted and on the received signals.  It  includes bit generation, LDPC coding, constellation mapper, MIMO precoding and pulse shaping. It allows also to program the RF front-end through USB interface, including beam steering settings and it is connected via Ethernet to the BEEcube platforms.
The BEEcubes are FPGA prototyping platforms equipped with four 3.52\,Gsamples/s  analog-to-digital converters (ADC) and digital-to-analog converters (DAC) and four FPGAs. 
BEEcubes are used to create the baseband waveform at the transmitter and to capture the baseband waveform at the receiver side. 
The  mm-wave  radios with beamforming capabilities are a 16 phased antennas mm-wave transmitter based on imec\rq{}s PHARA4 mm-wave radio chips   \cite{7417999}. 
These chips are direct-conversion transceivers with baseband phase shifting feature allowing analog beamforming functionality into the 57-66GHz frequency range. The resolution of the phase and amplitude
control of each antenna element is eight bits.

The MU-MIMO 60 GHz testbed is shown in Figure \ref{fig:testbed_picture}.
Two receiver units, constituted  by  imec's PHARA4 chips mounted on evaluation board, are allocated simultaneously.
The UEs use only one RF chain,  exploiting a  4 antenna phased array to perform analog beamforming.
A 32 antenna BS is implemented using  two 16-antennas modules. 
Hence, each RF chain has access to a single $2\times 8$  antenna array.
The antenna elements of every sub-array are integrated on the same board while different sub-arrays are separated by several wavelengths.

\section{Hybrid MIMO Measurement Results}
\label{sec:measurement}


In this section we present the measurement performed over the air in an indoor environment. 
The system uses QPSK modulation and LDPC coding with coding rate  fixed to $1/2$.
The transmitter and the two receivers are positioned at the same height  at 2.4\,m distance. The distance between the UEs is set to have an angular separation of 10$^\circ$ from the transmitter's perspective, making them very sensitive to inter-user interference.

\begin{figure}
\centering
\includegraphics[width=0.49\textwidth]{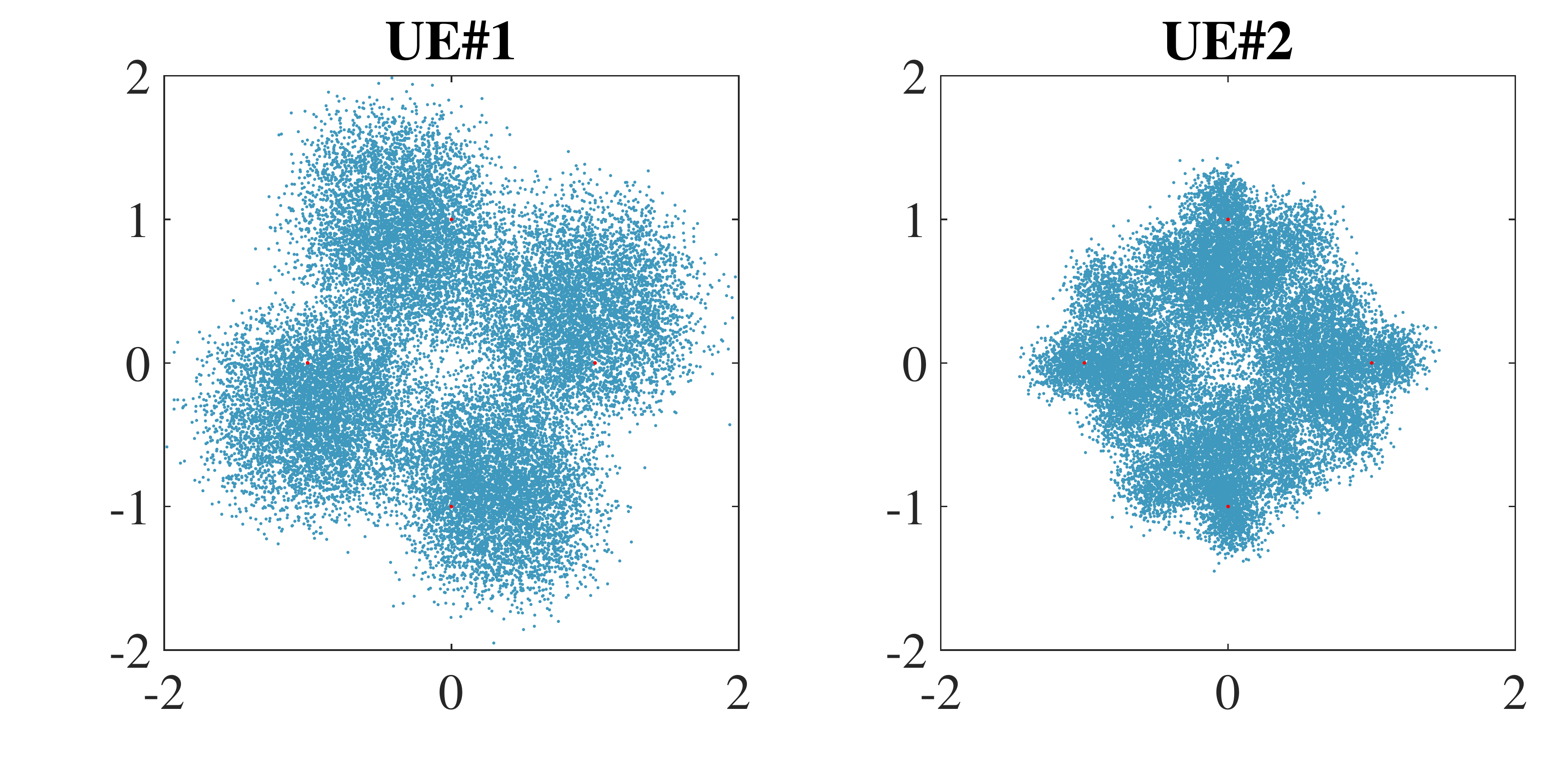}
 \caption{Received constellation analog beamforming. The two UEs are subjected to heavy interference.   \label{fig:analog}}
\end{figure}

In the first measurement we set the digital matrix $\mathbf{F}_D = \mathbf{I}$ per each subcarrier. The architecture reduces thus to the analog architecture presented in Figure \ref{fig:architectures}a in which the two sub-arrays operate independently, creating a point to point  connection with each UE using only analog processing.
The BS sets the beam to of each sub-array respectively to $\phi =-5^\circ, \theta = 0^\circ$ and $\phi =5^\circ, \theta = 0^\circ$  maximizing the received power at each UE.
Figure \ref{fig:analog} shows the constellation received by both UEs, where the impact of inter-user interference is visible, especially at UE\,2.
Despite a symmetrical scenario, UE\,2 receives a worst constellation. This can be attribute to the asymmetrical environment, which subjects the UEs to different multipaths propagation. Also  output power imbalances between the two RF chains can cause inequalities in the link.
\begin{table}[b!]
\caption{Measurement results.\label{tab:results}}
\begin{center}
  \begin{tabular}{l l l l}
    \hline
     & BER & PER & EVM [dB] \\ \hline  \hline
    Analog Beamforming UE 1 & $1.8\cdot10^{-2}$ & 0.81  &  -5.97\\ \hline
    Analog Beamforming UE 2 & $2.6\cdot10^{-2}$ & 0.73  &  -4.68\\ \hline
    Hybrid Beamforming UE 1 & $7.5\cdot10^{-4}$ & 0 &  -11.07  \\ \hline
    Hybrid Beamforming  UE 2 & $3.9\cdot10^{-4}$ & 0 &  -11.28  \\ \hline   
  \end{tabular}
\end{center}
\end{table} 
In Table\ref{tab:results} we report Bit Error Rate (BER), packet error rate (PER) and  error vector magnitude (EVM).
Even using a low modulation and coding scheme, user-interference is too heavy and fully analog systems cannot operate without errors.

In the second measurement, hybrid beamforming is used as depicted in Figure \ref{fig:architectures}b. The use of analog beamforming creates a $2\times2$  reduced MIMO channel.
Full reduced-channel estimation is used to design the RZF precoder. 
\begin{figure}
\centering
\includegraphics[width=0.49\textwidth]{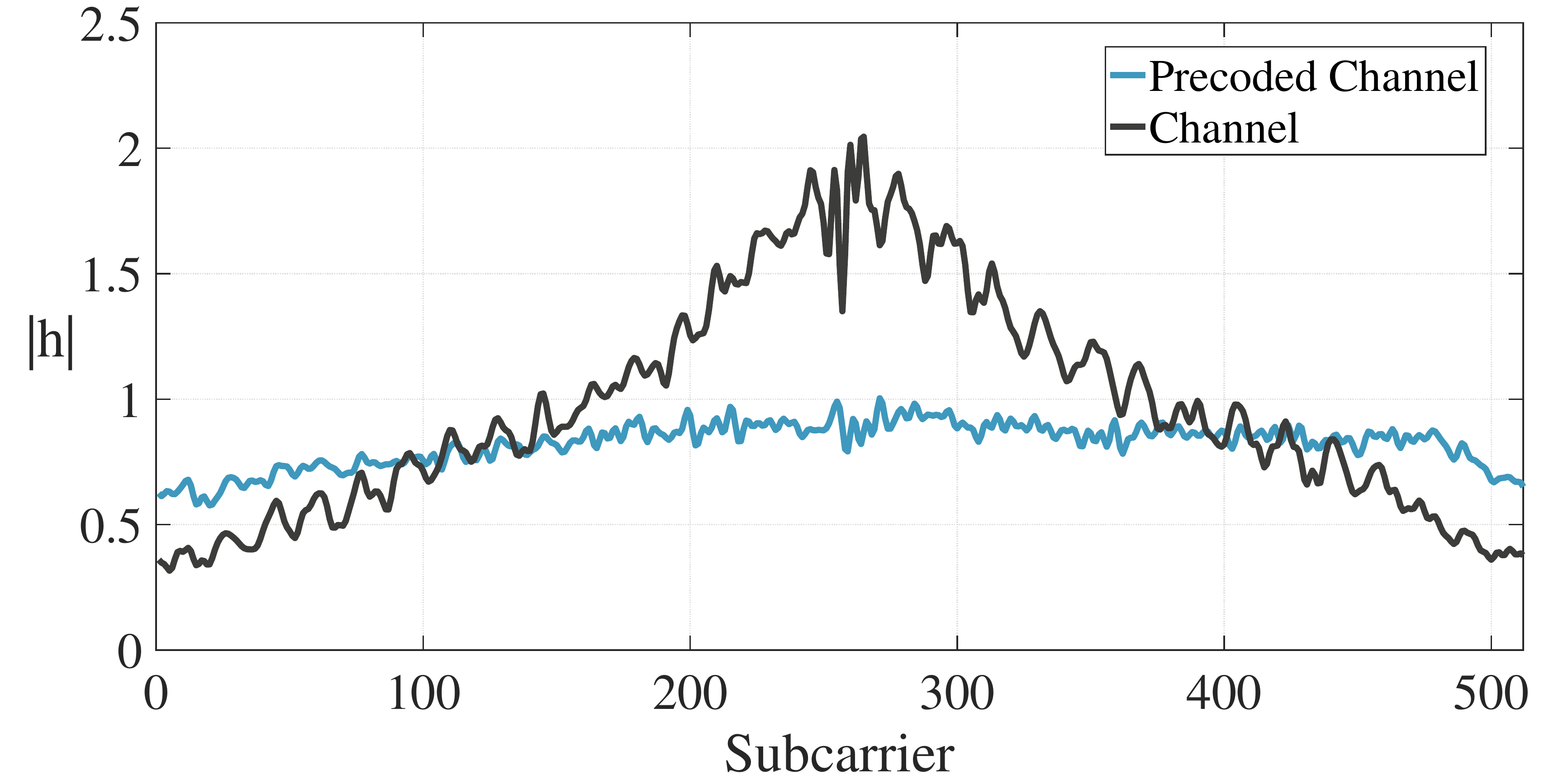}
 \caption{Frequency selective precoding compensates for channel impairments due to multipath propagation or transceiver imperfections.\label{fig:channel}}
\end{figure}
Figure \ref{fig:channel} shows both the non-precoded and the precoded version of the channel estimated by UE\,1 using respectively the CEF and the precoded CEF.  The non-precoded channel presents fluctuations caused by multi-path and a decay towards the edge of the bandwidth due to hardware bandwidth limitation. Frequency selective precoding at the transmitter lets the UE  experience  a flattened channel. 
\begin{figure}[t]
\centering
\includegraphics[width=0.49\textwidth]{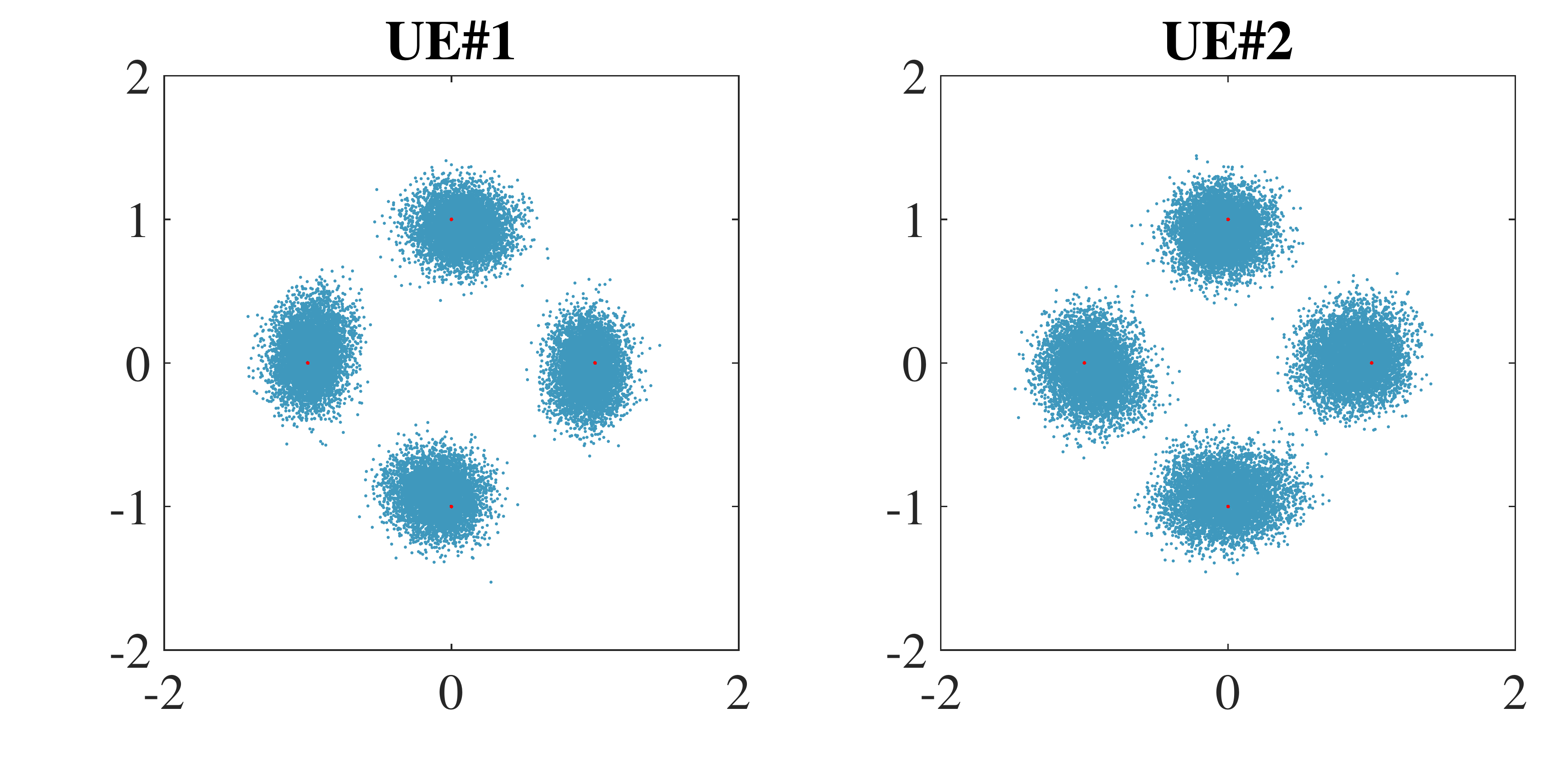}
 \caption{Received constellation hybrid beamforming. The RZF succeed in minimizing inter-user interference. \label{fig:hybrid}}
\end{figure}
Figure \ref{fig:hybrid} shows the received constellation by the two UEs which improve of around 6\,dB EVM compared a fully-analog architecture.
PER in this case is equal to zero for both the UEs which means that the hybrid scheme effectively succeeds in doubling the SISO throughput.

\section{Conclusion}
\label{sec_conclusion}

In this paper, we present a first realization of a multi-UE hybrid beamforming testbed including phased antenna arrays at both transmitter and receiver operating at 60\,GHz.
The base station equipped  with a 32 phased antenna array allocates simultaneously  two users and the frequency selective RZF succeed in  minimizing  the inter-user interference and it compensates the inevitable fluctuation of a wideband channel adapting the transmission along the frequency dimension.
Despite  the presence of channel multipaths, hardware non-idealities and analog beamforming imperfections,  hybrid MIMO systems guarantee spatial multiplexing even in interference-limited scenario, doubling the throughput compared to a SISO scenario and improving around 6\,dB the constellation EVM when compared to a fully analog architecture. 
Fully analog solutions cannot operate without errors due to interference and fully digital system are too costly and complex to realize. 
The results obtained suggests that hybrid MIMO is a good compromise to realize spatial multiplexing while keeping cost and complexity low. As a future work, there are still open questions to be addressed. The feasibility of hybrid MIMO in dynamic scenarios  need still to be proven. Moreover, testing a system with  more RF chains than UEs could shade light on the optimal hybrid configuration.

\bibliography{testbed}

\balance
\end{document}